# Novel depletion mode JFET based low static power complementary circuit technology

Artto Aurola[1], Vladislav Marochkin[2], and Mika Laiho[3]
[1]Hyperion Semiconductors Oy, Kadetintie 4 A 5, Helsinki, Finland, 00330
[2]Pixpolar Oy, Otakaari 5, Espoo, Finland, 02150
[3]Kovilta Oy, Piispanristintie 1, Piispanristi, Finland, 20760

Corresponding author: Artto Aurola (e-mail: artto.aurola@pixpolar.com).

**ABSTRACT** The lack of an easily realizable complementary circuit technology offering low static power consumption has been limiting the utilization of other semiconductor materials than silicon. In this publication, a novel depletion mode JFET based complementary circuit technology is presented and herein after referred to as Complementary Semiconductor (CS) circuit technology. The fact that JFETs are pure semiconductor devices, i.e. a carefully optimized Metal Oxide Semiconductor (MOS) gate stack is not required, facilitates the implementation of CS circuit technology to many semiconductor materials, like e.g. germanium and silicon carbide. Furthermore, when the CS circuit technology is idle there are neither conductive paths between nodes that are biased at different potentials nor forward biased p-n junctions and thus it enables low static power consumption. Moreover, the fact that the operation of depletion mode JFETs does not necessitate the incorporation of forward biased p-n junctions means that CS circuit technology is not limited to wide band-gap semiconductor materials, low temperatures, and/or low voltage spans. In this paper the operation of the CS logic is described and proven via simulations.

## I. INTRODUCTION

The Complementary Metal Oxide Semiconductor (CMOS) integrated circuit technology is at present the dominating integrated circuit technology in both analog and digital circuits. The reason for this is that it enables low power consumption, high operation speed, small foot print and low cost. A particularly advantageous feature of CMOS circuits is low static power consumption which stems from the fact that when CMOS circuits are idle there are neither conductive paths between nodes that are at different potentials nor forward biased p-n junctions.

The CMOS circuits are formed of n and p type enhancement mode Metal Oxide Semiconductor Field Effect Transistors (MOSFETs) comprising MOS stacks for realizing the gate and the channel of each MOSFET. The MOS stack comprises a thin insulator layer (e.g. oxide) sandwiched in between semiconductor and a conductor (e.g. metal) layers wherein the conductor layer forms the gate of the MOSFET.

The problem of CMOS circuit technology is that the MOS stack needs to be highly optimized so that it works reasonably well in both types of MOSFETs – a task that can be realized with ease only in silicon. In particular, the quality of the semiconductor insulator interface underneath the gate is of paramount importance for enabling the formation of both electron and hole inversion. The electron inversion layer forms the channel in the nMOSFET and the hole inversion layer forms the channel in the pMOSFET. Both electron and hole inversion can be realized relatively easily in silicon, but this is a significantly more difficult task in other semiconductor materials limiting their usability in integrated electronics.

The problem with chips made of other semiconductor materials than silicon has been so far that no low static power circuit technology has been available. Consequently, in systems including such chips auxiliary silicon-based chips are typically required comprising logic and/or analog circuitry for realizing necessary functionalities. This fact increases the number of components in the system, i.e., it lowers the integration level. Furthermore, the capacitance of a signal line on a circuit board interconnecting chips is much higher than the capacitance of a signal line on a chip. Consequently, the poorer the integration level the larger the power consumption and/or the lower the speed of the device. Moreover, the number of signal lines interconnecting chips is also limited reducing the programmability of lower integration level devices.

Wide-Band-Gap (WBG) semiconductor materials like silicon carbide (SiC) and gallium nitride (GaN) are utilized in many High Voltage (HV) devices because WBG semiconductors have a significantly higher voltage handling capability than silicon. However, HV devices utilizing WBG semiconductor chips comprise typically also auxiliary silicon chips incorporating complementary logic because at present there is no complementary circuit technology available that could be easily integrated on WBG semiconductor materials. Actually, the fact that the WBG semiconductor materials have a substantially larger band-gap than silicon results in both a significantly larger voltage handling capacity as well as substantially better tolerance to heat. Thus, the problem with auxiliary silicon chips is that they set the maximum operation temperature of the HV device and not the WBG semiconductor chips that could operate in a much higher temperature than silicon. Furthermore, the auxiliary silicon chips lower the integration level of HV devices. Consequently, the lack of integrated complementary logic in WBG semiconductor materials limits the operation temperature and reduces the programmability of HV devices.

Beside the HV devices many other applications would also benefit from monolithic chip arrangements based on other semiconductor materials than silicon, i.e., from System on Chip (SoC) circuits comprising complementary logic resources on the same non-silicon-based chip. For example, radiation detector arrangements incorporating pixelated non-silicon detector chips that are face to face bonded to silicon readout chips are problematic since in detector applications cooling is often required and the different temperature expansion coefficients of the different semiconductor materials may break either one of the chips. Furthermore, the face to face bonded chip arrangement reduces the yield of the system increasing the cost.

An interesting monolithic chip application would be germanium-based image sensors since they harvest much more ambient light than silicon-based image sensors during the darkest night time conditions. In particular, monolithic germanium-based image sensor chips incorporating the ultra-low-noise Modified Internal Gate (MIG) pixel technology [1] would enable unprecedented low light image quality due to the fact that in a MIG pixel interface generated 1/f noise and interface generated dark noise can be completely eliminated. The problem is, however, that it is difficult to realize working nMOSFETs in germanium-based CMOS due to lack of proper electron inversion [2]. Beside the image sensors, Light Emitting Diodes (LED) and lasers could be other interesting non-silicon-based monolithic chip applications.

Recently 2D semiconductor materials such as molybdenum disulfide have been intensively researched since in the future they could potentially act as a replacement for silicon in logic circuits. However, if CMOS based logic is applied on 2D semiconductor materials then the 2D materials require also a functional MOS stack, which adds up the complexity.

The fact that in a CMOS process a highly optimized MOS stack is required hinders also the implementation of CMOS logic in many silicon-based chips incorporating for example Micro Electro Mechanical Systems (MEMS) or power electronics.

Yet another problem in CMOS is that it is not trivial to realize on the same chip several different digital logic voltage spans like e.g. 1.8, 3.3 and 5 V due to the following reason. First of all, in order to realize the different voltage spans MOS stacks comprising different oxide/insulator thicknesses are typically required. Secondly, the oxide/insulator layers need to be very carefully manufactured and the manufacturing of each oxide/insulator layer involves high temperature steps that increase the diffusion of implanted impurity atoms. Consequently, the number of different digital logic voltage spans in CMOS chips is typically limited to two. However, it would be beneficial if one could generate a large number of different digital logic voltage spans on the same chip. The reason for this is that one could realize both timing and amplitude generation on the same chip instead of performing amplitude generation somewhere else on a Printed Circuit Board (PCB). This would improve the integration level, programmability, and speed as well as reduce the power consumption, cost, and design cycle of electronic devices.

Last but not least, ionizing radiation increases the amount of positive oxide charge in the oxide layer of the MOS stack shifting the threshold voltage of MOSFETs, which is particularly a problem for CMOS space applications. Furthermore, the ionizing radiation may also introduce irreparable damage to the MOS stack in CMOS integrated circuits leading to device failure.

As previously described, the MOS stack causes several problems for CMOS. In order to mitigate these problems, it is actually possible to get rid of the MOS stack by replacing the enhancement mode MOSFETs with similar type enhancement mode JFETs as has been done in [3]. The problem is, however, that in this manner forward biased p-n junctions are introduced in the circuitry increasing the static power consumption unless WBG semiconductor material and/or low temperature is utilized. Even if a wide band-gap semiconductor material is utilized the forward bias anyhow significantly limits the possible voltage span in the circuitry, which is a problem for analog applications.

In order to address afore described limitations of CMOS circuitries we have developed a novel complementary semiconductor circuit technology, referred to as Complementary Semiconductor (CS) circuit technology, that is compatible to CMOS manufacturing and operation. The basic idea is that one can replace any MOSFET in a CMOS circuit with two opposite type depletion mode JFETs. The CS circuit technology enables low static power consumption because during steady state there are neither forward biased p-n junctions nor conductive paths between nodes that are biased at different potentials. Although two JFETs consume more space and are likely slower than one MOSFET they provide other benefits.

## II. JFET characteristics

The proposed CS circuit technology relies in the use of depletion mode JFETs. The benefit of JFETs is that they can be easily realized with implantations [4, 5], i.e., a highly optimized MOS stack enabling both electron and hole inversion is not required. Thus, CS circuits are significantly easier to manufacture than CMOS circuits and can be implemented virtually on any semiconductor material provided that a) large enough single crystal wafers are available, b) both n and p type impurity atoms having shallow enough intra-band-gap energy states are available, c) the diffusion constants of the n and p type impurity atoms are such that annealing can be performed without altering the concentration profile of the implanted impurity atoms too severely, d) annealing can be performed without outgazing of elements corresponding to the semiconductor material and/or to the n and p type impurity atoms, and e) low resistance contacts can be made to both n and p type doped semiconductor material.

The threshold voltage of JFETs can be easily changed by simply adjusting the channel implant dose [6] (i.e., there is no need to tamper a carefully optimized MOS stack). The ability to adjust the threshold voltage is beneficial in both analog and digital circuits; the channel implant dose accurately defines the threshold voltage of JFETs unlike in MOSFETs wherein the threshold voltage is affected by the amount of positive oxide charge which is not trivial to control. Moreover, the JFET channel is less noisy than the MOSFET channel due to lack of interface induced noise [6, 7]. These points are advantages particularly in analog applications.

Furthermore, the dopant profile in JFET channel (and thus also the threshold voltage of JFETs) remains fixed for long periods of time even in high flux ionizing radiation environments; JFET operation is neither affected by build-up of positive oxide charge like MOSFET operation nor by degradation of minority carrier life time like Bipolar Junction Transistor (BJT) operation. The build-up of positive oxide charge and degradation of minority carrier life time are naturally occurring phenomena in high flux ionizing radiation environments. Due to aforesaid reasons JFET-based circuits offer significant benefits in space applications.

The fact that JFETs can be easily manufactured via implants means that they can also be easily integrated to a CMOS process, i.e., the addition of new implantation steps does not alter the CMOS process – it only increases the number of implantation mask steps required in the process. Actually, it is possible to replace any part of a CMOS circuitry with a CS circuitry resulting in a hybrid circuit corresponding to both CS and CMOS circuit technology.

Figure 1a corresponds to a schematic layout of a pJFET that can be utilized in a CS circuit. The black dashed lines in figure 1a correspond to cross-sections presented in figures 1b, 1c, and 1d. The abbreviation STI refers to Shallow Trench Isolation, i.e., to insulator trenches that in present case reach down to a buried insulator layer. Dark blue corresponds to p+ type source/drain dopings (source on the left, drain on the right), light blue corresponds to a p type channel doping, light red corresponds to an n well doping forming a back gate, and bright red corresponds to n+ type dopings forming a front gate (located in between the source and the drain) as well as an n well contact doping (located on far left). The front gate doping is shallower than the n well contact doping to enable a shallower JFET design. In figure 1a the horizontal black dashed line corresponds to the cross-section presented in figure 1b, the vertical black dashed line on the left runs along the front gate and corresponds to the cross-section presented in figure 1c, and the vertical black dashed line on the right runs in between the front gate and the drain and corresponds to the cross-section presented in figure 1d. It is important to note that in the pJFET of figures 1a – d, the front gate is connected via the back-gate doping to the back-gate contact doping and thus these dopings are always at the same potential.

Figure 2 corresponds to a schematic cross-section of an nJFET that can be utilized in a CS circuit. In figure 2 bright red refers to n+ type source/drain dopings (source on the left, drain on the right), light red refers to an n type channel doping, light blue corresponds to a p well doping forming a back gate, dark blue refers to p+ type dopings forming a front gate (located in between the source and drain) as well as a p well contact doping (located on far left). The nJFET is structurally similar to the pJFET of figure 1 except that the p and n type doped regions are interchanged. Thus, other cross-sections and the layout of the nJFET are not shown. Furthermore, the front gate is biased by the p well contact doping via the back-gate contact doping.

## II. Novel CS inverter

In order to demonstrate how CS circuit technology [8] can be applied in practice we concentrate the investigation on the inverter, which is the most fundamental digital circuit component. In order to facilitate the description of the CS inverter, the CMOS inverter is discussed first. The benefit of the CMOS inverter (and CMOS circuit technology in general) is that during steady state there are a) neither forward biased p-n junctions b) nor conducting paths between nodes that are biased at different potentials, i.e., during steady state virtually no current is flowing from a node biased e.g. at 5 V to a node biased at 0 V. Thus, the steady state power dissipation of the CMOS inverter is low.

The CMOS inverter comprises one n type and one p type MOSFET as is presented in the schematic drawing of figure 3a – the symbol of the CMOS inverter is presented in figure 3b. The operation of the CMOS inverter is described next. The source of the nMOSFET is connected to 0 V, the source of the pMOSFET is connected e.g. to 5 V, the gates of both MOSFETs are connected together as input, and the drains of both MOSFETs are connected together as output. When 0 V is connected to the input the channel of the nMOSFET is nonconductive and the channel of the pMOSFET is conductive resulting in 5 V bias at the output. On the other hand, when 5 V is connected to the input the channel of the nMOSFET is conductive and the channel of the pMOSFET is nonconductive resulting in 0 V bias at the output.

A schematic drawing of the CS inverter configuration is presented in figure 4a and the corresponding symbol in figure 4b. In the CS inverter configuration of figure 4a the nMOSFET of the CMOS inverter of figure 3a is replaced with a plate capacitor, static potential sources providing -1.5 and 0 V, p type depletion mode JFET (referred to as J1 or input pJFET), and n type depletion mode JFET (referred to as J2 or output nJFET). Correspondingly, the pMOSFET of the CMOS inverter of figure 3a is replaced with following items: plate capacitor, static potential sources providing 0 and 6.5 V, n type depletion mode JFET (referred to as J3 or input nJFET), and p type depletion mode JFET (referred to as J4 or output pJFET). In figure 4a a number (0.3 or 1.8) is associated to all of the JFETs and this number corresponds to the channel pinch-off voltage. The channel pinch-off voltage in both output nJFET and output pJFET is 0.3 V and the channel pinch-off voltage in both input pJFET and input nJFET is 1.8 V.

In this publication the channel pinch-off voltage of a depletion mode transistor refers to the reverse bias between the gate and the source when the channel transforms from non-depleted to depleted, i.e., below the channel pinch-off voltage a non-depleted area extends from the source to the channel. When the channel is non-depleted the channel is conductive if temperature is above the dopant freezeout temperature. It should be also noted that a) the channel pinch-of voltage is less dependent on temperature than the threshold voltage, b) the channel pinch-off voltage refers to reverse bias, which is defined in this publication as a positive quantity, c) in a depletion mode transistor the absolute value of the threshold voltage is smaller than the channel pinch-off voltage, and d) the value of the channel pinch-off voltage would be negative in an enhancement mode transistor (i.e. the pinch-off voltage would refer to forward bias between gate and source).

The drains of the output nJFET and output pJFET are connected together as the output of the CS inverter configuration and the gates of the input pJFET and input nJFET are connected together as the input of the CS inverter configuration. The source of the input pJFET is connected to -1.5 V, the source of the input nJFET is connected to 6.5 V, the source of the output nJFET is connected to 0 V, and the source of the output pJFET is connected to 5 V. It is important to note that during steady state the potential on the input and output can be either 0 or 5 V, i.e. these potentials correspond to the binary logic potential pair in the CS binary logic circuitry.

Furthermore, the potentials -1.5 V and 6.5 V do not correspond to logic potentials.

The p type drain of the input pJFET and the p type gate of the output nJFET are connected together and this entity is referred to as the P type Internal Node (PIN). The PIN is coupled to the input via a plate capacitor, which is later on referred to as the PIN capacitor. The n type drain of the input nJFET and the n type gate of the output pJFET are connected together and this entity is referred to as the N type Internal Node (NIN). The NIN is coupled to the input via a plate capacitor, which is later on referred to as the NIN capacitor.

In all JFETs of the inverter configuration of figure 4a the front gate and the back gate are connected together, which is a beneficial but not a necessary arrangement. Furthermore, the PIN and NIN plate capacitors are also beneficial but not mandatory. The role of the plate capacitor is to act as a source of constant capacitance since the gate to source capacitance in a JFET depends significantly on the gate to source potential, i.e., with the help of the plate capacitor the capacitance of the gate node remains significant also when a large reverse bias is applied between the gate and the source.

It should be noted that all the potentials -1.5, 0, 5, and 6.5 can be separately connected to the chip or alternatively one can connect only one potential pair like e.g. -1.5 and 6.5 V or 0 and 5 V to the chip. If e.g. 0 and 5 V are connected to the chip, then it is possible to generated -1.5 and 6.5 V on-chip e.g. with appropriate charge pumps.

*A. DESCRIPTION OF OPERATION*
The CS inverter configuration of figure 4a is operated in the following manner. The input of the inverter is connected to either one of the binary logic potential levels, namely, 0 and 5 V. When the input is set at 0 V the channel of the input pJFET is non-depleted because the reverse bias between the gate and source is 1.5 V, which is below the 1.8 V channel pinch-off voltage of the input pJFET, i.e., the channel of the input pJFET is clearly conductive. Thus, the drain of the input pJFET and the gate of the output nJFET are set to -1.5 V, i.e. the PIN is set to -1.5 V. Since the channel pinch-off voltage of the output nJFET is 0.3 V, the gate to source reverse bias is 1.2 V above the channel pinch-off voltage and thus the channel of the output nJFET is nonconductive virtually at any temperature. Due to the nonconductive channel the drain potential of the output nJFET is not defined by the source of the output nJFET.

Furthermore, when the input is set at 0 V the channel of the input nJFET is depleted because the reverse bias between the gate and source is 6.5 V, which is 4.7 V above the 1.8 V channel pinch-off voltage of the input nJFET. Thus, the channel of the input nJFET is nonconductive at any temperature. The important point is, however, that when the input is set from 5 to 0 V then the input drags the NIN along via the NIN capacitor forcing the NIN to settle to the source potential of the output pJFET, i.e., the NIN is set to 5 V. The channel pinch-off voltage of the output pJFET is 0.3 V and thus the gate to source potential is 0.3 V below the channel pinch-off voltage meaning that the channel of the output pJFET is non-depleted and thus conductive. Consequently, the output of the inverter configuration is set to 5 V.

When the input of the inverter configuration is set to 5 V the channel of the input nJFET is non-depleted because the reverse bias between the gate and source is 1.5 V, which is below the 1.8 V channel pinch-off voltage of the input nJFET, i.e., the channel of the input nJFET is clearly conductive. Thus, the drain of the input nJFET and the gate of the output pJFET are set to 6.5 V, i.e. the NIN is set to 6.5 V. Since the channel pinch-off voltage of the output pJFET is 0.3 V, the gate to source reverse bias is 1.2 V above the channel pinch-off voltage and thus the channel of the output pJFET is nonconductive virtually at any temperature. Due to the nonconductive channel the drain potential of the output pJFET is not defined by the source of the output pJFET.

Moreover, when the input is set at 5 V the channel of the input pJFET is depleted because the reverse bias between the gate and source is 6.5 V, which is 4.7 V above the 1.8 V channel pinch-off voltage of the input pJFET. Thus, the channel of the input pJFET is nonconductive at any temperature. The important point is that when the input is set from 0 V to 5 V then the input drags the PIN along via the PIN capacitor forcing the PIN to settle to the source potential of the output nJFET, i.e., the PIN is set to 0 V. The channel pinch-off voltage of the nJFET is 0.3 V and thus the gate to source potential is 0.3 V below the channel pinch-off voltage meaning that the channel of the output nJFET is non-depleted and thus conductive. Consequently, the output of the inverter configuration is set to 0 V.

### B. LAYOUT OF A CS INVERTER AND CORRESPONDING IMPLANTS

A possible layout of the CS inverter configuration of figure 4a is presented in figure 5 comprising pJFETs and nJFETs corresponding to figures 1 and 2. In all JFET layouts the lengths of back gate contact doping, source, front gate, drain as well as the distance between each of them is chosen to be 0.6 m. The channel widths of the input pJFET, input nJFET, and output nJFET are the same, but the channel width of the output pJFET is twice as large. Correspondingly, the NIN capacitance is also twice as large as the PIN capacitance in order to compensate the larger channel width of the output pJFET. The reason for the larger channel width of the output pJFET is that it compensates the fact that the mobility of holes is smaller than the mobility of electrons enabling thus more symmetric inverter operation. For the same reason in CMOS circuits the width of the pMOSFETs is typically larger than the width of the nMOSFETs. The reason why the input pJFET is not wider is that it is not necessary since the input pJFET drives a much smaller capacitance than the output pJFET. It is important to note that also the NIN capacitor is double the size when compared to the PIN capacitor.

A CS inverter corresponding to the arrangement of figures 4a and 5 is designed by selecting proper layout dimensions as well as proper implantation energies and doses. The length of the front gate was chosen to be 0.6 μm in all JFETs. The substrate material in all four JFETs is n type silicon having a resistivity of 2000 Ωcm. In both pJFETs the n type front gate corresponds to an arsenic implant 1e14 $cm^{-2}$, 50 keV, 0° tilt (dose, energy, tilt) and the n type back gate corresponds to a phosphorus implant 5e13 $cm^{-2}$, 800 keV, 0° tilt. In both nJFETs the p type front gate corresponds to a BF2 implant 1e14 $cm^{-2}$, 20 keV, 0° tilt and the p type back gate corresponds to a boron implant 2.25e14 $cm^{-2}$, 480 keV, 0° tilt. In the input pJFET the p type channel corresponds to a boron implant 3.7e12 $cm^{-2}$, 90 keV, 0° tilt and in the output pJFET the p type channel doping corresponds to a boron implant 2.25e12 $cm^{-2}$, 90 keV, 0° tilt. In the input nJFET the n type channel corresponds to a phosphorus implant 7.5e12 $cm^{-2}$, 270 keV, 0° tilt and in the output nJFET the n type channel corresponds to a phosphorus implant 5e12 $cm^{-2}$, 270 keV, 0° tilt. Furthermore, the implants undergo standard thermal process steps including e.g. annealing. The only requirement for the n and p type contact implants is low contact resistance.

### III. Simulation results of the CS inverter

#### A. STATIC TWO-DIMENSIONAL SIMULATION

The JFETs and the corresponding manufacturing processes as described in II.2 are implemented to the process and device physics Synopsys TCAD simulator in order to generate two-dimensional (2D) cross-sections of the JFETs. The 2D cross-sections are further implemented to the simulated CS inverter configuration comprising an output pJFET that is two times wider than the other JFETs, necessary wirings, and PIN and NIN plate capacitors. The NIN plate capacitor is also twice as large as the PIN plate capacitor in order to comply with the two times wider output pJFET. In addition, an output plate capacitor is added between the output node and the ground. However, as far as static simulations are concerned the capacitance values of the PIN, NIN and output plate capacitors have only very limited effect, if any, on the simulation results.

Figure 6a corresponds to a static two-dimensional potential simulation of the CS inverter wherein the sources of the JFETs are configured to the appropriate potentials (-1.5, 0, 5, 6.5 V) and the input is biased at 0 V (i.e. at the first logic potential). The four potential curves in figure 6b correspond to potentials graphs on vertical cut-lines (indicated by black dashed lines) that run along the middle of the front gates of each JFET in figure 6a. In figure 6a the white lines correspond to the border of the depletion region. When the input is at 0 V one can deduce based on the depletion region borders that the channels of both pJFETs are non-depleted and thus conductive and that the channels of both nJFETs are depleted. The fact that the channel of the input pJFET is conductive sets the PIN to -1.5 V and the fact that the channel of output pJFET is conductive sets the output to 5 V.

When the input is at 0 V one can deduce based on figures 6a and 6b also the following simulation results. First, in the output nJFET the potential barrier for electrons from the source to the drain is around 1.3 V. Second, in the input nJFET the potential barrier for electrons from the drain to the source is around 3.1 V. Due to afore said high barriers the channels of both nJFETs are nonconductive when the input is at 0 V.

Figure 7a corresponds to a static two-dimensional potential simulation of the CS inverter wherein the sources of the JFETs are configured to the appropriate potentials (-1.5, 0, 5, 6.5 V) and the input is biased at 5 V (i.e. at the second logic potential). The four potential curves in figure 7b correspond to potentials on vertical cut-lines (indicated by black dashed lines) that run in the middle of the front gates of each JFET in figure 7a. When the input is at 5 V one can deduce based on the depletion region borders that the channels of both nJFETs are non-depleted and thus conductive and that the channels of both pJFETs are depleted and thus nonconductive. The fact that the channel of the input nJFET is conductive sets the NIN to 6.5 V and the fact that the channel of output nJFET is conductive sets the output to 0 V.

When the input is at 5 V one can deduce based on figures 7a and 7b the following simulation results. First, in the output pJFET the potential barrier for holes from the source to the drain is around 1.2 V. Second, in the input pJFET the potential barrier for holes from the drain to the source is around 2.9 V. Due to afore said high barriers the channels of both pJFETs are nonconductive when the input is at 5 V.

It is important to note that in order to comply with figure 4a, the channel doses in the input JFETs have been adjusted such that the channels become roughly depleted when 1.8 V reverse bias is applied between the gate and the source/drain dopings. Correspondingly, the channel doses in the output JFETs are adjusted such that the channels become roughly depleted when 0.3 V reverse bias is applied between the gate and the source/drain dopings.

### A. TRANSIENT TWO-DIMENSIONAL SIMULATION

In this simulation study the plate capacitance per channel width Cp was chosen to be same for both PIN and NIN, i.e., the PIN and NIN capacitors were not optimized separately. However, the size of the NIN capacitance was set to be twice as large than the size of the PIN capacitance since the output pJFET is two times wider than the output nJFET. Furthermore, the size of the output plate capacitor Cout was set to be four times larger than the maximum input capacitance of the inverter. The reason for the latter point is that the ultimate purpose of these transient simulations is to emulate a CS inverter in a Fan-Out four (FO4) arrangement, wherein the output of the inverter is connected to the inputs of four similar inverters and the input is connected to a four times smaller inverter. An important point is, however, that the transient simulations in this publication do not fully correspond to the FO4 arrangement since the input is connected to an almost ideal clock signal and since the capacitance of the output plate capacitor is fixed and corresponds to four times the value of maximum input capacitance throughout the voltage span between 0 and 5 V.

It should be also noted, that the introduction of PIN and NIN plate capacitors could be avoided by appropriately fine tuning the inherent diffusion capacitances, but this would require significant work, i.e., it is a much simpler approach to optimize the size of the corresponding plate capacitor Cp. The maximum capacitance of the input without the PIN and NIN plate capacitors was simulated to be 2.6 fF/μm. Thus, in the simulations of the CS inverter corresponding to the approximate FO4 arrangement the following equation

$$C_{out} = 10.4 \, fF/\mu m + 12 C_p \qquad (1)$$

was used to represent the output capacitance $C_{out}$. In figures 8a and 8b different values for the plate capacitor $C_p$ are implemented while 0.1 ns rise and fall times for the input signal are used. After the rise/fall the input stays fixed at 0 or 5 V for 9.9 ns after which the input is changed again. It can be deduced from figures 8a and 8b that the most optimal value for $C_p$ is 1.5 fF/μm corresponding to almost symmetrical output fall and rise times. The corresponding $C_{out}$ was therefore 28.4 fF/μm.

In this paper, the propagation delay is defined as the time difference between the moment when the input has reached 50% of the transition (i.e. 2.5 V) and the moment when the output has reached 50% of the transition (i.e. 2.5 V). Based on figure 8a, the propagation delay of the simulated CS inverter in the approximate FO4 configuration is 1.7 ns when the output nJFET drives the output and 1.9 ns when the output pJFET drives the output. Correspondingly, in the simulated CS inverter operated in the approximate FO4 configuration, the time it takes for the output to change from 10% to 90% level is 3.4 ns when the output nJFET drives the output (i.e. output potential changes from 4.5 to 0.5 V) and 3.5 ns when the output pJFET drives the output (i.e. output potential changes from 0.5 to 4.5 V).

Figure 9a represents a transient two-dimensional potential simulation result of the CS inverter in the approximate FO4 configuration corresponding to the results presented in figure 8a just before the input is changed from 0 to 5 V and wherein $C_p$ is 1.5 fF/μm. Figure 9b represents potential graphs on vertical cut-lines (indicated by black dashed lines) that are running along the centers of the front gates in figure 9a. By comparing the transient simulation result of figure 9b with the static simulation result of figure 6b (in both cases input is at 0 V) one can deduce that the only major difference is that in the transient simulation the gate of the output pJFET (being part of NIN) is still forward biased with respect to the source. The benefit of the forward bias is that it enhances significantly the conductivity of the p type channel in the output pJFET and thus speeds up the transient when input is changed from 5 to 0 V.

Figure 10a represents a transient two-dimensional potential simulation result of the CS inverter in the approximate FO4

configuration corresponding to figure 8a just before the input is changed from 5 to 0 V and wherein Cp is 1.5 fF/μm. Figure 10b represents potential graphs on vertical cut-lines (indicated by black dashed lines) that are running along the centers of the front gates in figure 10a. By comparing the transient simulation result of figure 10b with the static simulation result of figure 7b (in both cases input is at 5 V) one can deduce that the only major difference is that in the transient simulation the gate of the output nJFET (being part of PIN) is still forward biased with respect to the source. The benefit of the forward bias is that it enhances significantly the conductivity of the n type channel in the output nJFET and thus speeds up the transient when input is changed from 0 to 5 V.

Transient simulation results on the approximate FO4 arrangement are represented in figure 11a showing the potentials at the PIN, NIN, input, and output during several clock cycles. It should be noted that the simulation results during the first 10 ns differ from the results after this period due to the fact that the first 10 ns corresponds to a steady continuation of a static initial state. Based on these simulation results, it is clear, that only in the transient stage PIN is significantly forward biased with respect to the source of the output nJFET when output nJFET is conductive. Similarly, only in the transient stage NIN is significantly forward biased with respect to the source of the pJFET when output pJFET is conductive. A close-up of the initial static phase and the first transient cycle is shown in figure 11b.

One can observe from figures 11a and 11b that the capacitive coupling between the input and the NIN and PIN causes the potential on NIN and PIN to be dragged along when the potential at the input is changed. Moreover, when the input potential is changed and NIN and PIN are not yet biased via the channel of the corresponding input JFET then the potential on PIN is driven first to be more negative than -1.5 V and the potential on NIN is driven first to be more positive than 6.5 V. However, the PIN settles via current running in the channel of the input pJFET to -1.5 V in roughly 2 ns and the NIN settles via current running in the channel of the input nJFET to 6.5 V in roughly 1 ns. Furthermore, one can observe from figures 11a and 11b that when input is at 5 V, i.e. when PIN is not biased via the channel of the input pJFET, then PIN is at least 0.5 V forward biased with respect to the source of the output nJFET. Correspondingly, when input is at 0 V, i.e. when NIN is not biased via the channel of the input nJFET, then NIN is at least 0.3 V forward biased with respect to the source of the output pJFET.

Transient simulation results on the approximate FO4 arrangement showing the source currents in input pJFET and nJFET as well as in output pJFET and nJFET are presented in figure 12a and a close-up of figure 12a is shown in figure 12b. Based on the graphs one can deduce that the output JFETs are never simultaneously conductive meaning that there is no short circuit power dissipation in the simulated CS inverter. In addition, one can deduce that source currents are far larger in the output JFETs than in the input JFETs.

In figure 13 transient simulation results on the approximate FO4 arrangement represent source currents in the output JFETs as well as output potential. The current at the output is simply the sum of the source currents of both output JFETs. Based on figure 13 it is clear that both the output potential and the output current settle down well during a 20 ns cycle time, i.e. when 50 MHz clock rate is used. Moreover, based on figure 13 it should be even possible to reduce the cycle time to 10 ns corresponding to 100 MHz clock rate.

## IV. LEVEL SHIFTING OPERATION IN A CS INVERTER

Figure 14a corresponds to a schematic drawing of a CS inverter configuration wherein the output binary logic pair is shifted with respect to the input binary logic pair. In this case, the channel pinch-off voltages of the input JFETs are altered when compared to the inverter configuration of figure 4a by simply adjusting the channel dopings of the input JFETs. In the inverter of figure 14a the p type channel doping of the input pJFET is increased so that the channel pinch-off voltage is shifted from 1.8 V to 2.8 V when compared to the inverter of figure 4a. Correspondingly, the n type channel doping of the input nJFET is reduced so that the channel pinch-off voltage is shifted from 1.8 V to 0.8 V. In addition, the source potential of the input pJFET is changed from -1.5 V to -2.5 V and the source potential of the input nJFET is changed from 6.5 V to 5.5 V. No physical changes are made to the output JFETs, but the source potential of the output nJFET is changed from 0 V to -1 V and the source potential of the output pJFET is changed from 5 V to 4 V. In this manner, the output binary logic potential levels (-1 V & 4 V) are shifted by -1 V with respect to the input binary logic potential levels (0 V & 5 V). Figure 14b corresponds to the symbolic representation of the inverter configuration of figure 14a.

Even larger shifts in binary logic levels can be realized by simply connecting CS inverters in series. In figure 15 there are 6 CS inverters connected in series wherein in each of the inverters the logic levels are shifted by -1 V. So, in the chain of 6 series connected CS inverters the output logic levels are shifted altogether by -6 V when compared to the input logic levels (the initial input binary logic pair being 0 and 5 V and the eventual output binary logic pair being -6 and -1 V). Because there is an even number of inverters connected in series the output of the inverter chain corresponds to -6 V when the input is at 0 V and the output of the inverter chain corresponds to -1 V when the input is at 5 V.

## V. CS NAND CIRCUIT

A two input CMOS NAND cell is presented in figure 16a and a corresponding two input CS NAND cell is presented in figure 16b wherein each MOSFET of figure 16a is replaced with two opposite type depletion mode JFETs and a plate capacitor. It is important to note that one can convert any CMOS circuit in a similar manner to a CS circuit, i.e., by replacing one MOSFET with two opposite type depletion mode JFETs and one optional plate capacitor.

## VI. DISCUSSION AND FUTURE WORK

The inverter was an example aiming to show operation of the CS circuit technology qualitatively. However, in order to be able to characterize the CS circuit technology more rigorously it would be necessary to investigate a SoC based on the CS circuit technology. This is, however, beyond the scope of this publication. Moreover, further efforts are needed to reduce the JFET area, the capacitances, and the voltage swing in CS circuit technology. Furthermore, it would be interesting to investigate the fundamental limit of scaling of the JFET area and the voltage swing in order determine the minimum achievable cost and power consumption of the CS circuit technology.

The CMOS circuit technology offers many benefits like mature manufacturing, small footprint, low cost, small capacitance, low voltage, small dynamic power consumption, and high clock rate. An important aspect is that there are many applications wherein the cost and power consumption are not fundamentally important parameters, but clock rate is. Yet another point is that it may be impossible to compensate a low clock rate by any other means. For this reason, couple of possible ways are discussed below how one could be able to improve the clock rate in CS circuits.

The first obvious way to increase the clock rate in CS circuits is naturally to scale down the JFET area and to reduce the voltage swing. The former scales down the parasitic capacitances and the latter one scales down the time needed to charge the capacitances. These points are naturally also beneficial from the cost and power consumption point of view.

A second possibility to increase the speed of CS logic could be to modify the channel doping of JFETs in such a manner that in the middle of the channel there would be a heavily doped thin stripe in between two almost intrinsic regions of the same doping type. Alternatively, one could modify the channel doping of JFETs in CS logic in such a manner that there would be two heavily doped thin stripes of the same doping type surrounded by almost intrinsic regions of the same doping type. The benefit of the almost intrinsic regions in the channel would be that the mobility of such low doped regions is very high. In particular, the almost intrinsic region between the two heavily doped stripes could turn out to be very useful.

A third possibility to increase the speed of CS logic could be to utilize a multi-channel depletion mode JFET architecture. A possible layout of a multi-channel nJFET is presented in figure 17a and corresponding cross-sections in figures 17b and 17c. Although only 3 stacked channels are illustrated in figure 17 the number of the stacked channels could be naturally much larger than that.

A fourth possibility to increase the speed of CS logic could be to utilize output JFETs incorporating the following features:
- when the gate is at the same potential than the source then the channel would be conductive but depleted,
- when input is changed from one logic potential to another then in the output JFET driving the output the gate to source forward bias would turn the channel non-depleted, and
- the decay time of afore said forward bias between the source and gate would be such that the channel would be again depleted when half of the cycle time is reached, i.e., before the input is changed again.

The benefit of such output JFET arrangement is that the gate to source capacitance in an output JFET that is turned from the conductive stage to the nonconductive stage is small throughout the switching process due to the depleted channel. On the other hand, the capacitance of the output JFET that is turned from the nonconductive stage to the conductive stage is large throughout almost the entire switching process due to the non-depleted and thereby highly conductive channel. Consequently, during switching the input capacitance and the corresponding input current would mainly feed the output current and the part of the input current responsible for switching off the other output JFET would be relatively small. This aspect should also reduce the propagation delay.

A fifth possibility to increase the speed of CS logic would be to minimize the input capacitances. The input capacitance could be minimized e.g. by configuring the CS logic in such a manner that the channels of all JFETs would be depleted but conductive when the gate is at same potential than the source. In this manner the capacitances in the JFETs would be very small and thus one could utilize also very small PIN and NIN plate capacitors. The conductive but depleted channels are naturally not as conductive as non-depleted channels, but this fact could be compensated by utilizing the multi-channel JFET architecture as well as by implementing first and second logic potentials that are closely spaced.

It is important to note that it is naturally possible to combine afore said different possibilities to increase the speed of CS logic. In figure 18 a signal booster arrangement is presented comprising two CS inverters and having 0 and 1 V as logic potentials at the input and output. The first CS inverter on the left could be for example designed according to afore said fifth possibility and the second inverter on the right could be for example designed according to afore said fourth possibility. Since in this manner the input capacitance of the signal booster would be minimized also the RC time constant of the input line would be minimized. On the other hand, the large potential swing (5 V) on the node between the two inverters should provide a large current on the output of the signal booster. It should be investigated whether such signal booster arrangement could provide sufficiently large current amplification with reasonably small delay and footprint.

## VII. CONCLUSION

In this paper, a depletion mode JFET based novel low static power complementary circuit technology is introduced and the operation is proven via simulations. We anticipate that this CS circuit technology would be particularly well suited for space and analog applications as well as for implementing integrated logic to different semiconductor materials than silicon e.g. in order to provide Internet of Things (IoT) chips made of wide band-gap semiconductor materials having very low static power consumption without posing prohibitive limitations to the voltage span and operation temperature. Furthermore, it should be easier to integrate CS than CMOS circuitry to some silicon-based technologies like e.g. MEMS chips. We foresee also that the CS circuit technology could be integrated to CMOS logic chips in order to facilitate the generation of I/O nodes that enable programmable amplitudes for I/O signals.

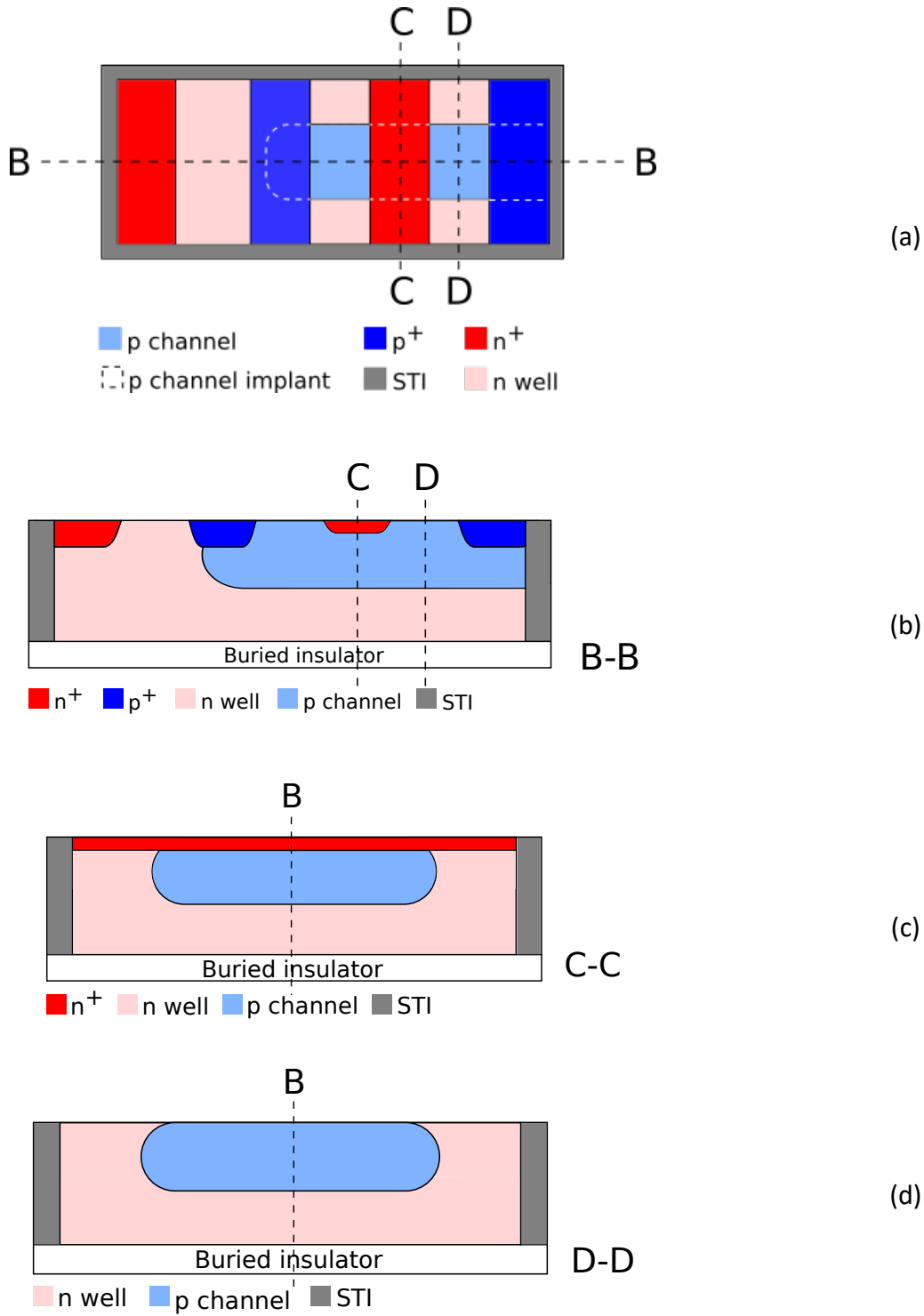

Figure 1. Schematic drawings of a pJFET configuration that can be utilized in the CS inverter configuration: (a) Layout of the pJFET, (b) horizontal cross-section of the pJFET, (c) vertical cross-section of the pJFET along the front gate, (d) vertical cross-section of the pJFET taken in between the front gate and the drain.

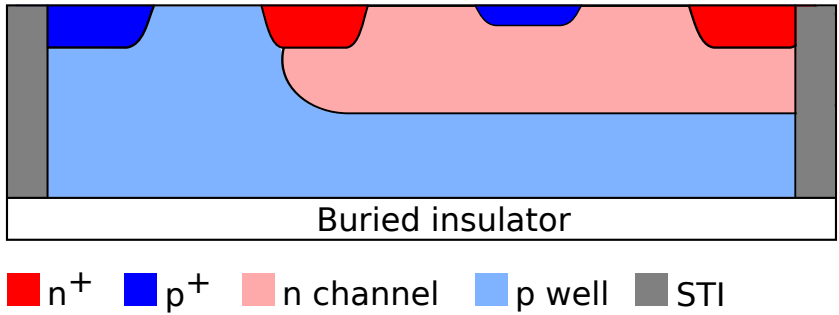

Figure 2. A schematic cross-section of an nJFET that can be utilized in the CS inverter configuration.

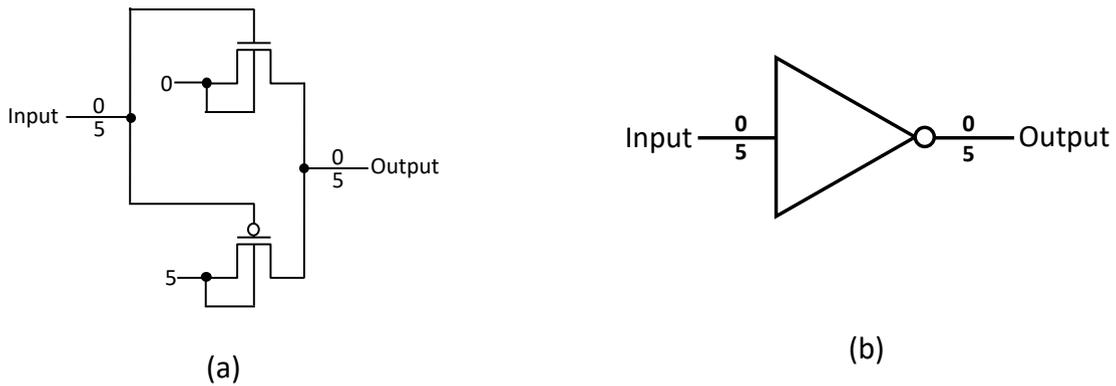

Figure 3. (a) Schematic drawing of a CMOS inverter comprising n and p type enhancement mode MOSFETs, (b) Symbol of the CMOS inverter.

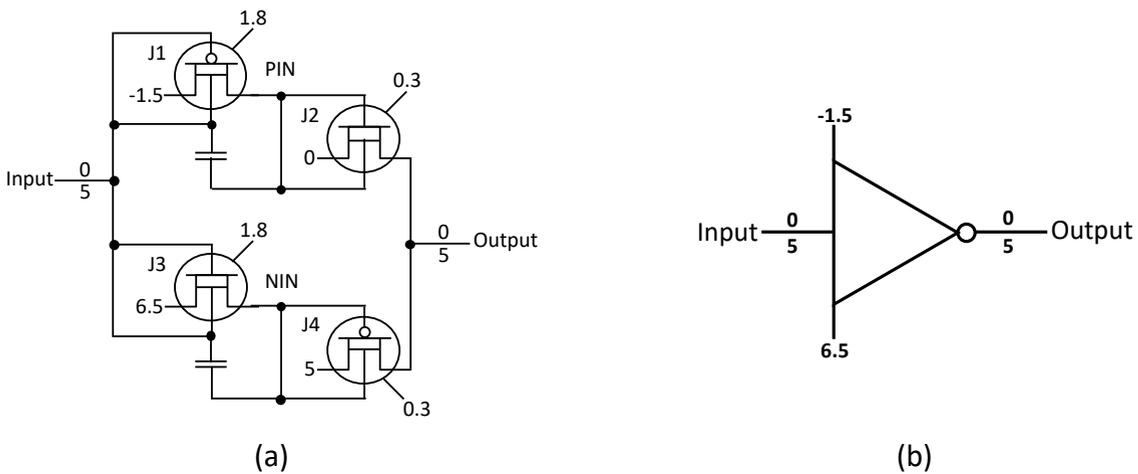

Figure 4. (a) Schematic drawing of a CS inverter configuration comprising two n type and two p type depletion mode JFETs (the channel pinch-off voltages 0.3 and 1.8 V are also indicated), two capacitors, and four static potential sources: -1.5, 0, 5, and 6.5 V. The potentials 0 and 5 V correspond to the binary logic potential pair. (b) Symbol of the CS inverter configuration presented in figure 4a.

Figure 5. Schematic layout of the CS inverter configuration used in the simulations.

Figure 6a. Static two-dimensional simulation of electric potential in a CS inverter configuration corresponding to figures 1, 2, 4, and 5. When 0 V is applied to the input the output is set to 5 V. The white lines correspond to the depletion region boundaries. The channels in the input and output pJFETs are non-depleted and conductive, but the channels in the input and output nJFETs are depleted and nonconductive.

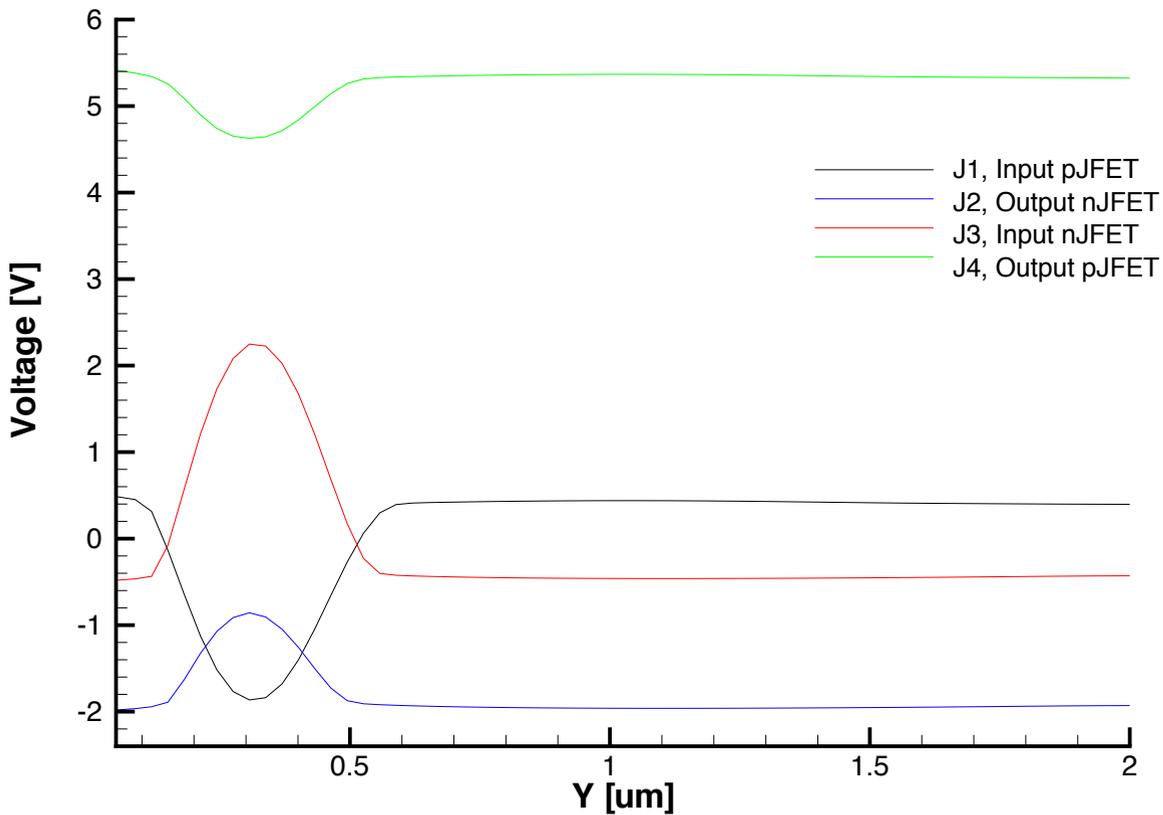

Figure 6b. Electrical potential curves on vertical cut-lines indicated by the black dashed lines in figure 6a. The vertical cut-lines run along the middle of the front gates.

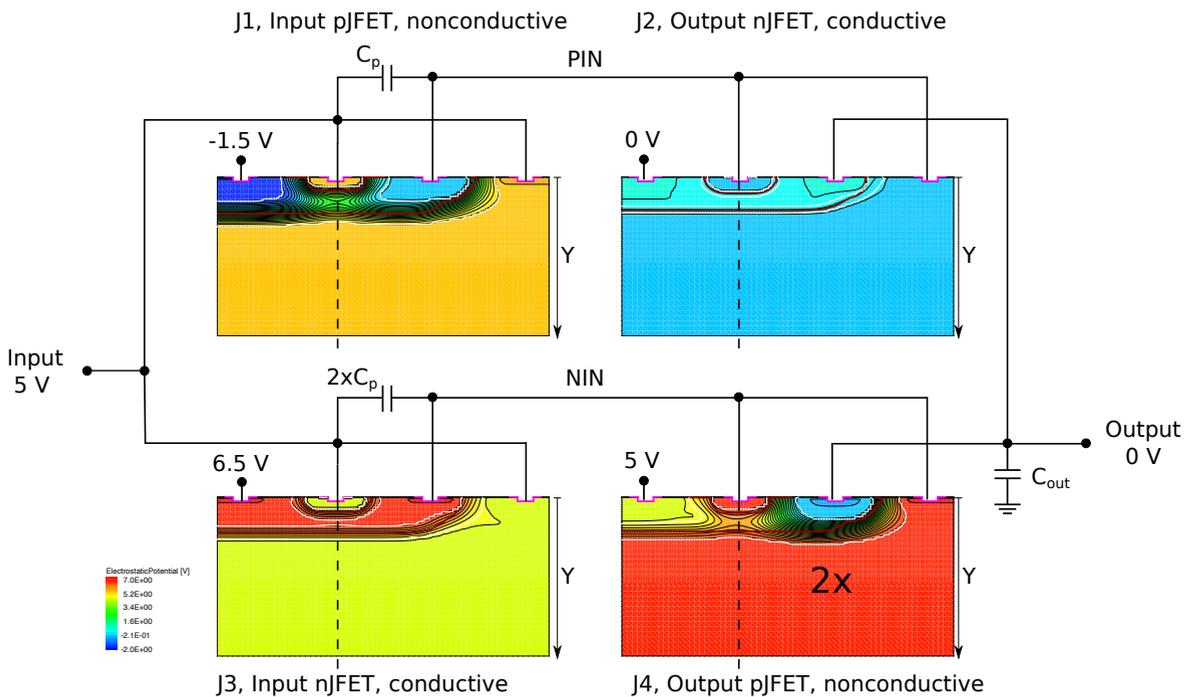

Figure 7a. Static two-dimensional simulation of electric potential in a CS inverter configuration corresponding to figures 1, 2, 4, and 5. When 5 V is applied to the input the output is set to 0 V. The white lines correspond to the depletion region boundaries. The channels in the input and output pJFETs

are depleted and nonconductive, but the channels in the input and output nJFETs are non-depleted and conductive.

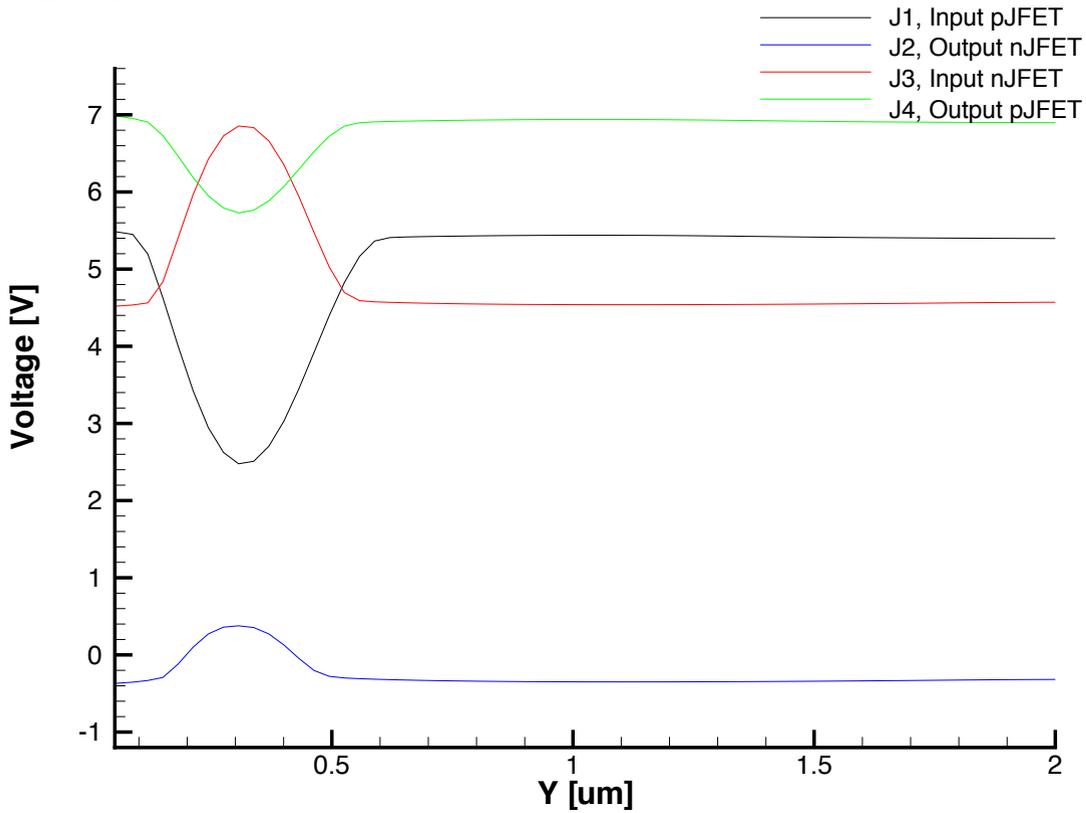

Figure 7b. Electrical potential curves on vertical cut-lines indicated by black dashed lines in figure 7a. The vertical cut-lines run along the middle of the front gates.

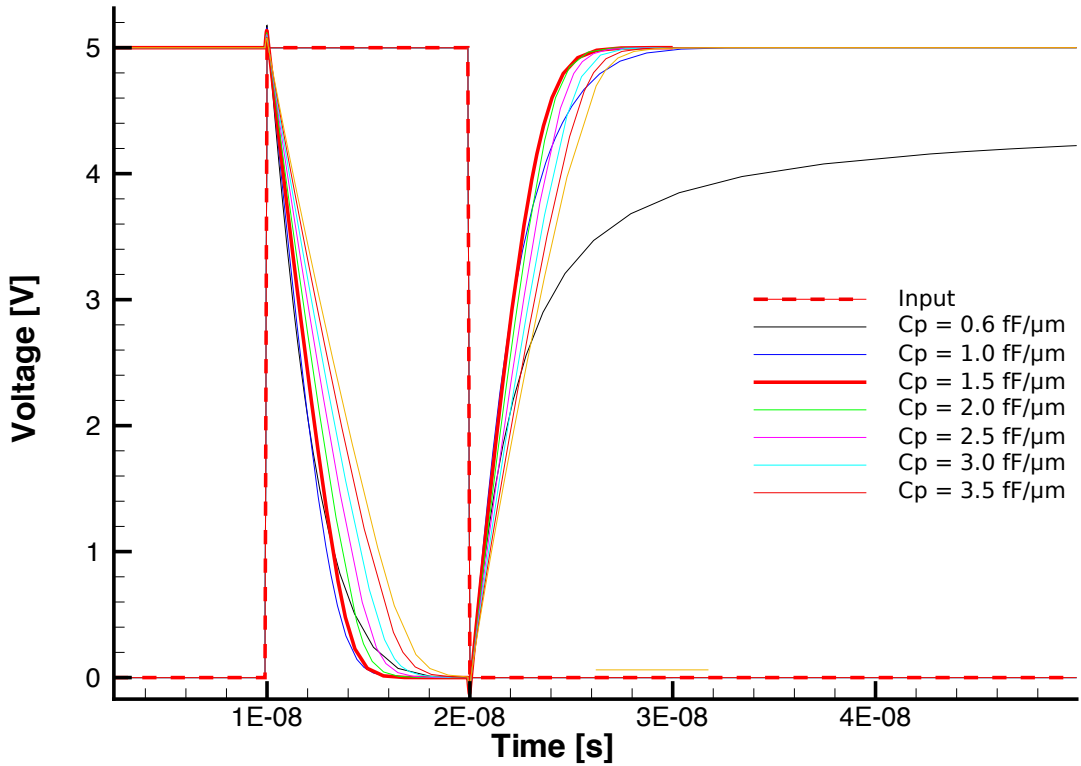

Figure 8a. The dashed line represents the predefined input potential versus time in the simulated CS inverter and the solid lines represent the corresponding responses on the output for various plate capacitor Cp capacitance values.

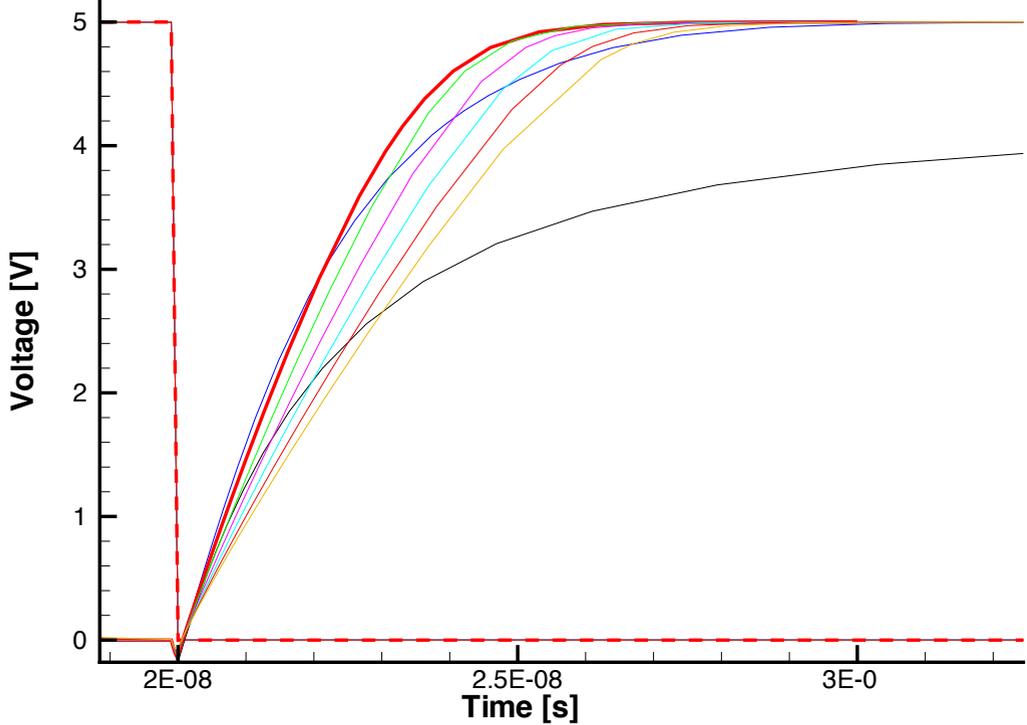

Figure 8b. Part of the figure 8a zoomed.

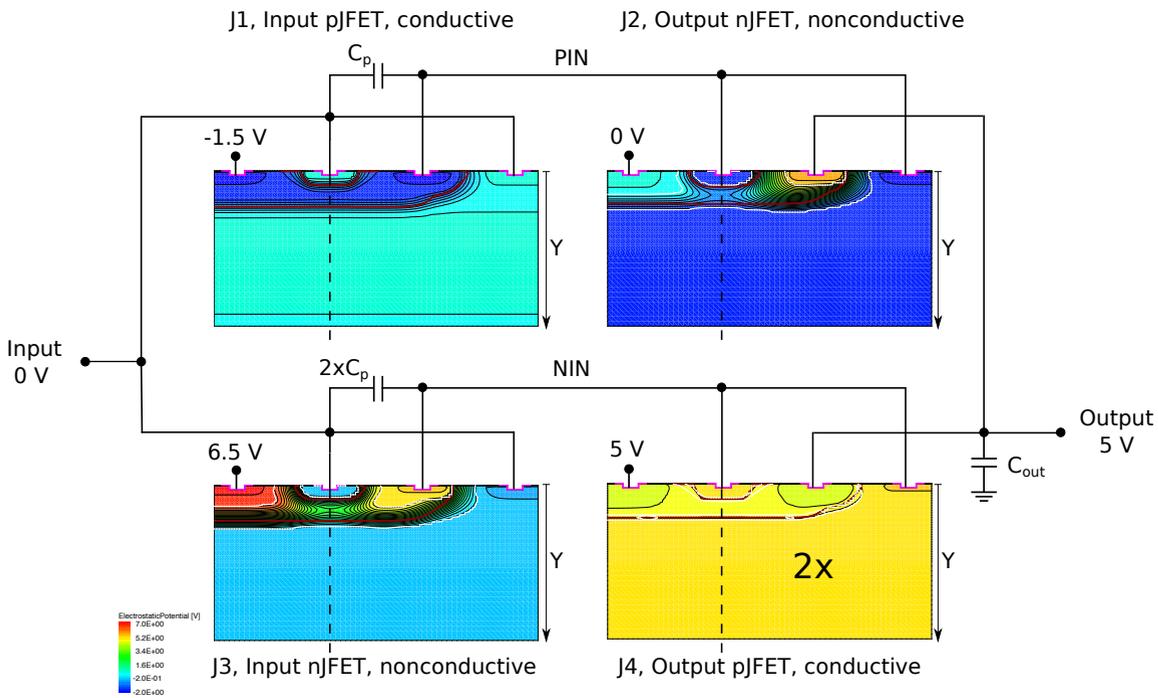

Figure 9a. Transient two-dimensional simulation of the electric potential in a CS inverter roughly 10 ns after the input is set from 5 to 0 V. At this time point the output is already settled to 5 V. The channels in the input and output pJFETs are non-depleted and conductive, but the channels in the input and output nJFETs are depleted and nonconductive. The gate of the output pJFET (part of NIN) is still forward biased with respect to the source of the output pJFET.

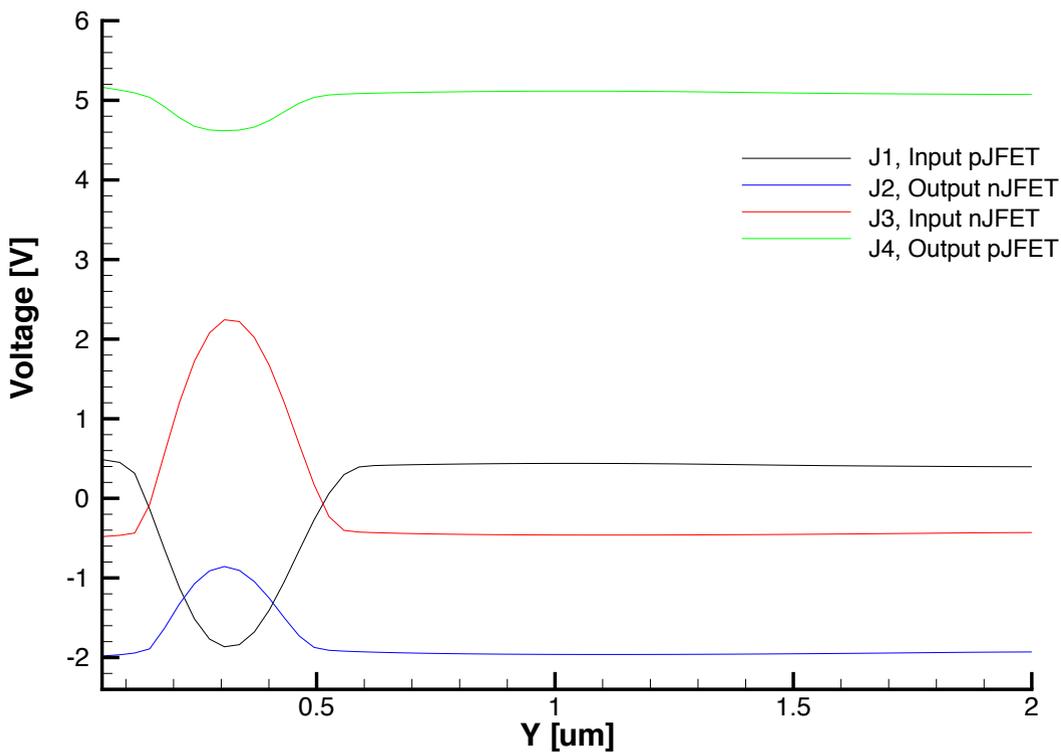

Figure 9b. Electrical potential curves on vertical cut-lines indicated by black dashed lines in figure 9a taken along the middle of the front gates. By comparing figure 9b with figure 6b it is clear that in case of figure 9b the gate of the output pJFET is still forward biased with respect to the source.

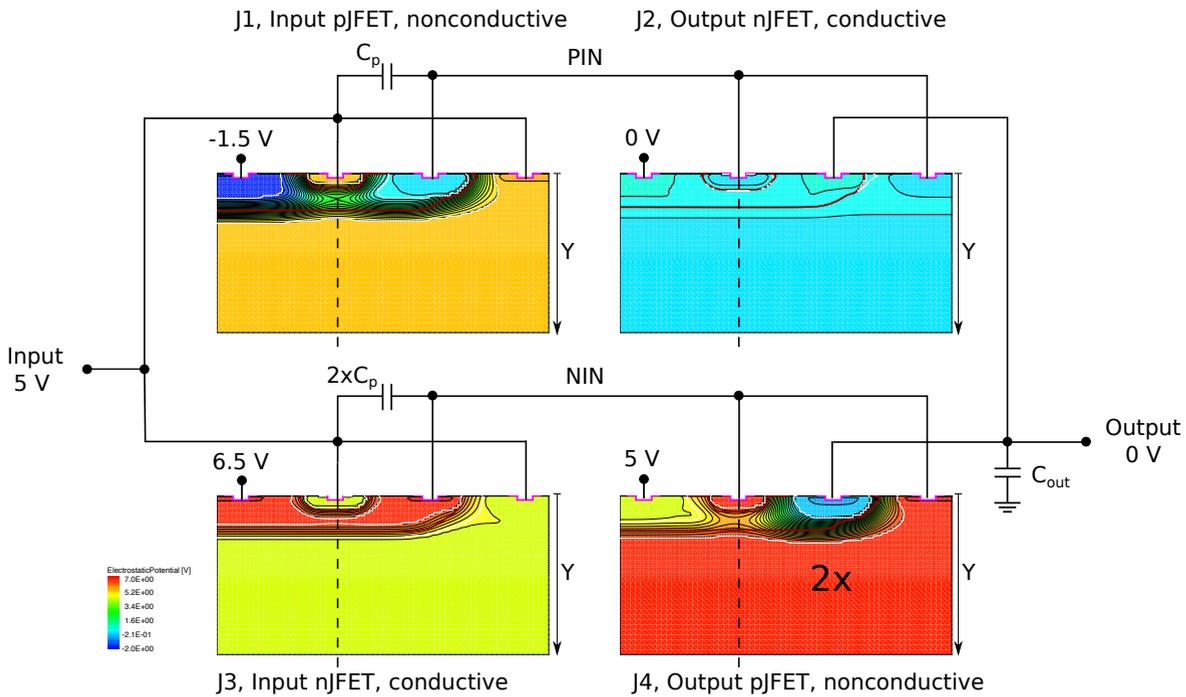

Figure 10a. Transient two-dimensional simulation of electric potential in a CS inverter roughly 10 ns after the input is set from 0 to 5 V. At this time point the output is already settled to 0 V. The channels in the input and output nJFETs are non-depleted and conductive, but the channels in the input and output pJFETs are depleted and nonconductive. The gate of the output nJFET (part of PIN) is still forward biased with respect to the source of the output nJFET.

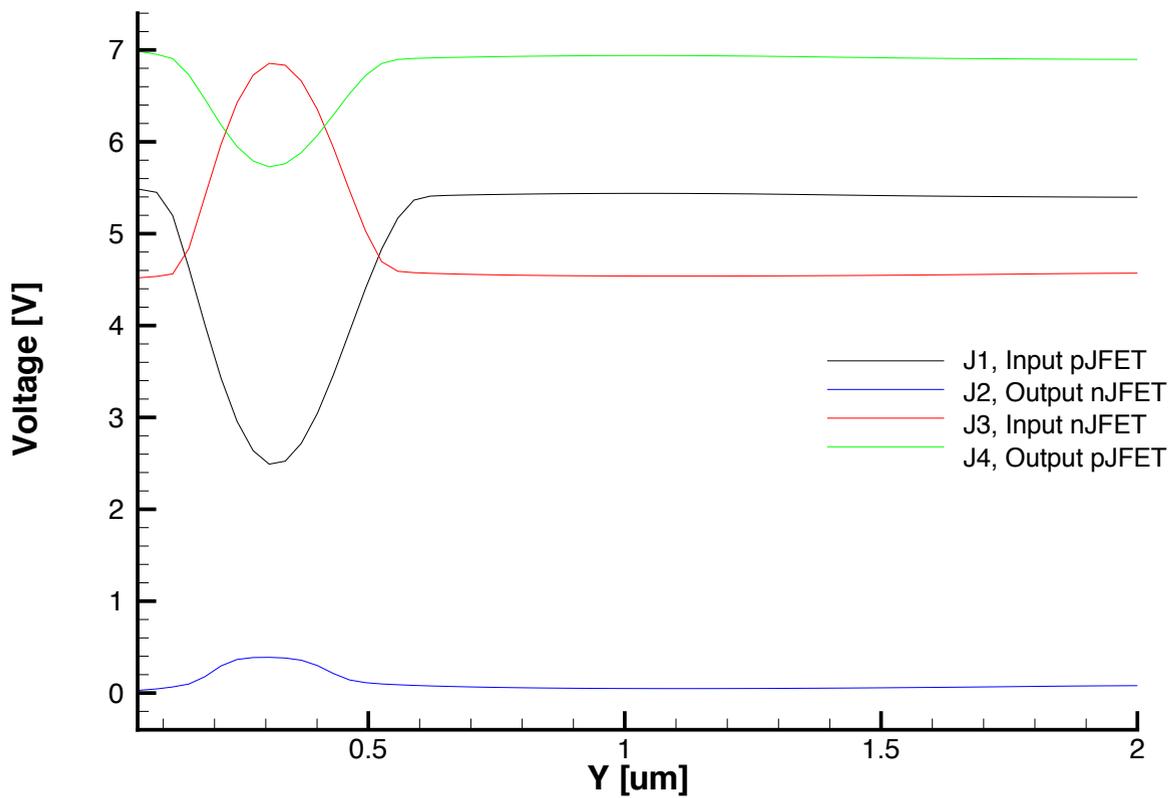

Figure 10b. Electrical potential curves on vertical cut-lines indicated by dashed black lines in figure 10a taken along the middle of the front gates. By comparing figure 10b with figure 7b it is clear that in case of figure 10b the gate of the output nJFET is still forward biased with respect to the source.

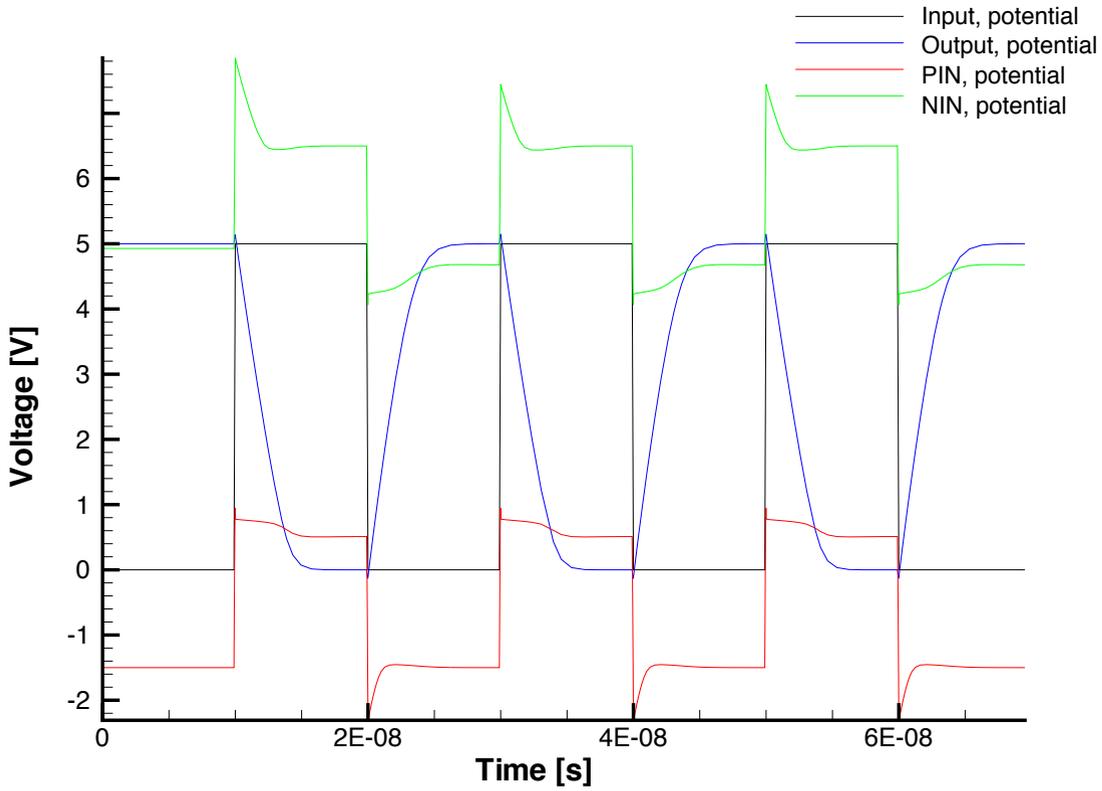

Figure 11a. Potentials versus time at input, output, PIN, and NIN in the simulated CS inverter.

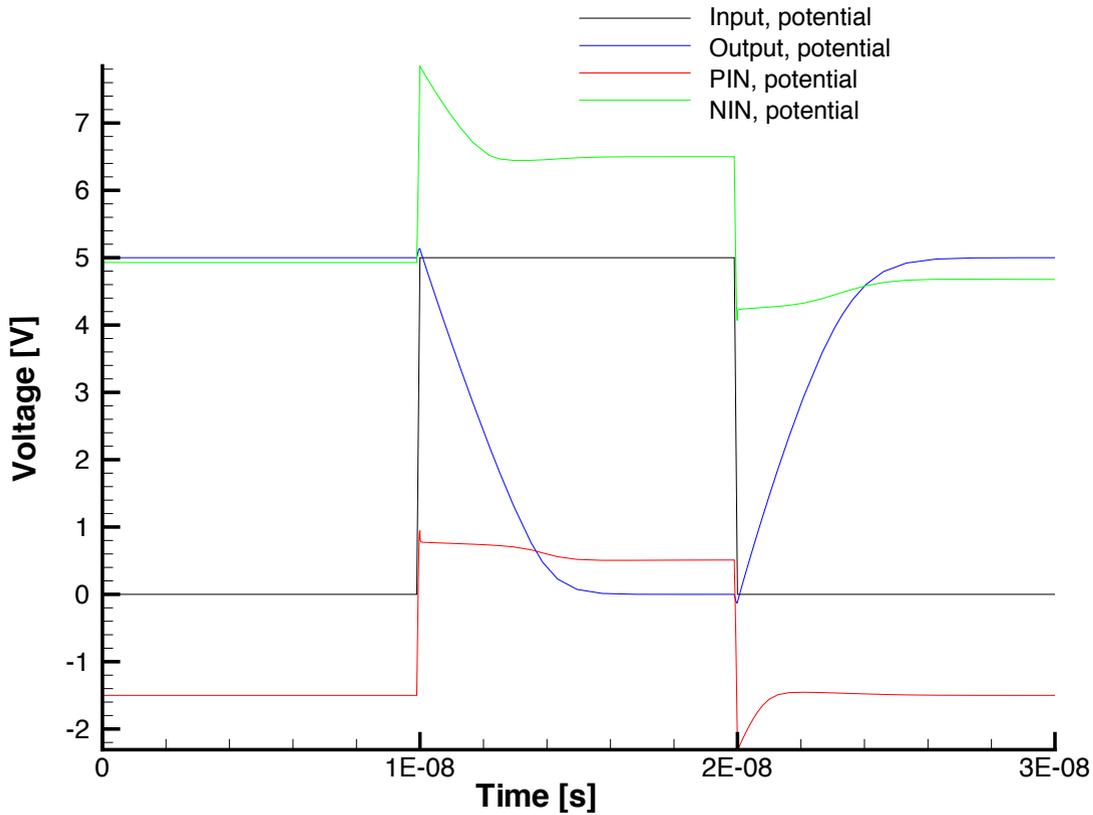

Figure 11b. Part of figure 11a zoomed.

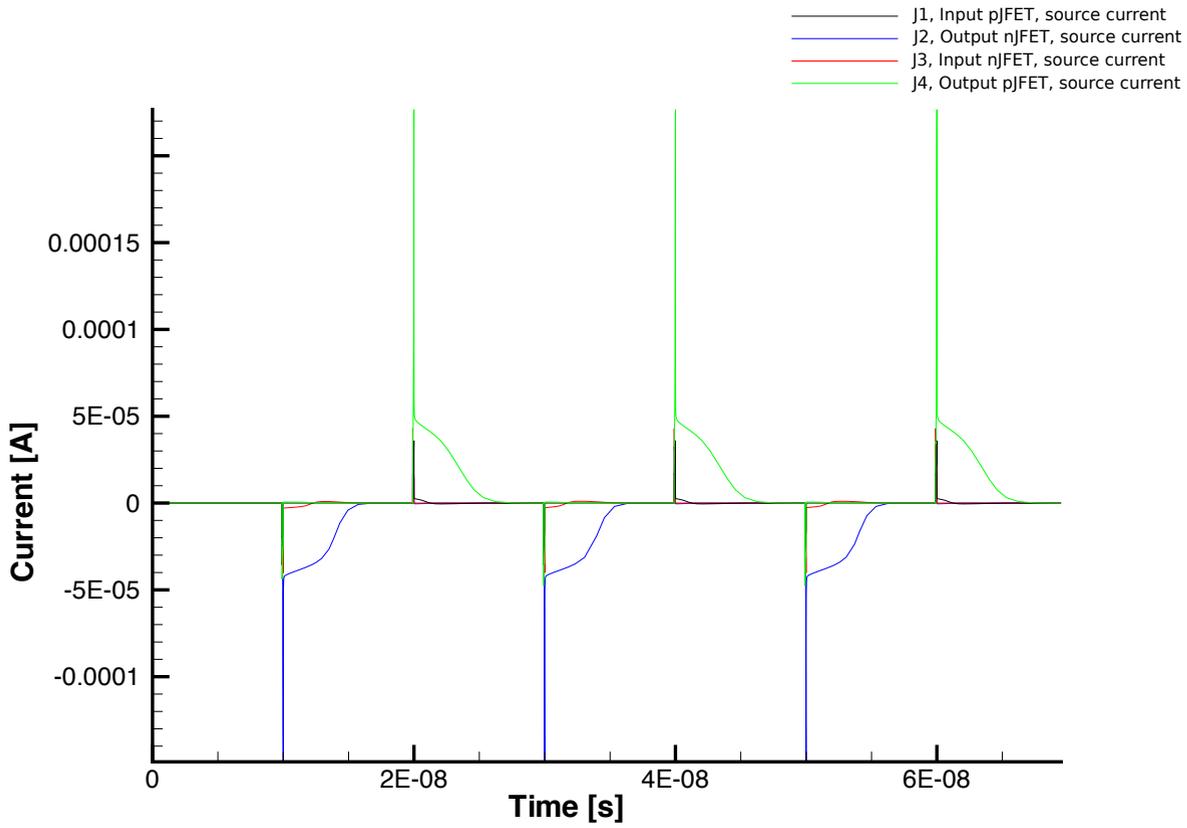

Figure 12a. Simulated source current versus time in all JFETs of the CS inverter.

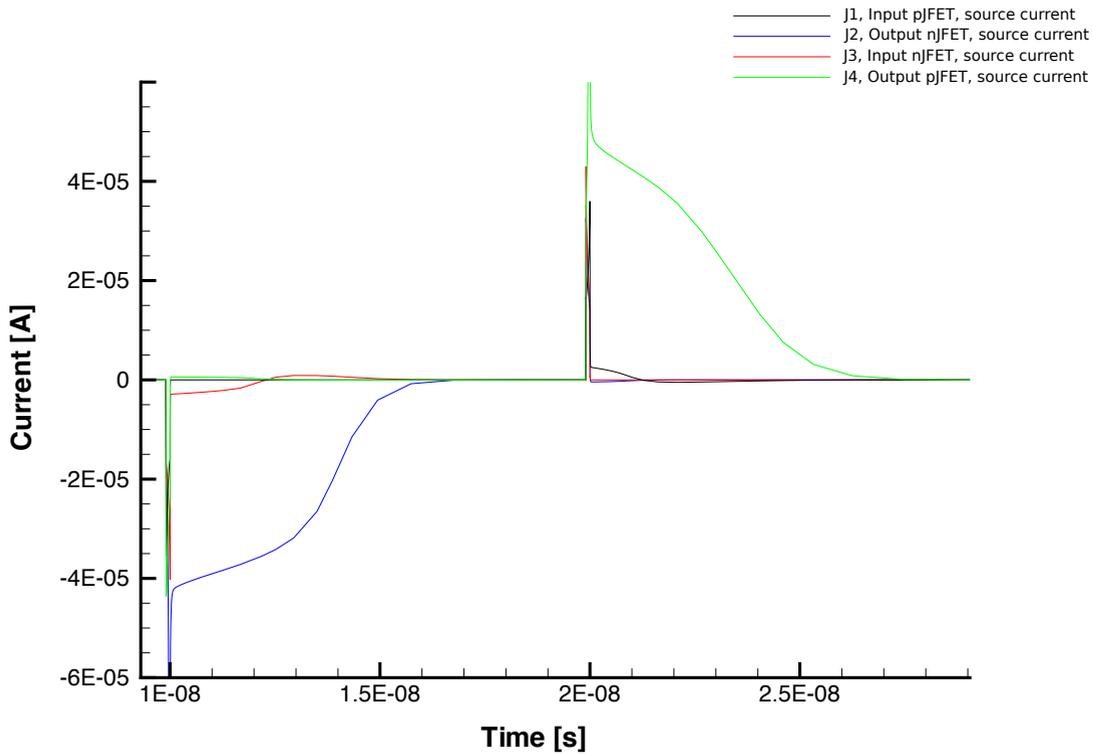

Figure 12b. Part of figure 12a zoomed.

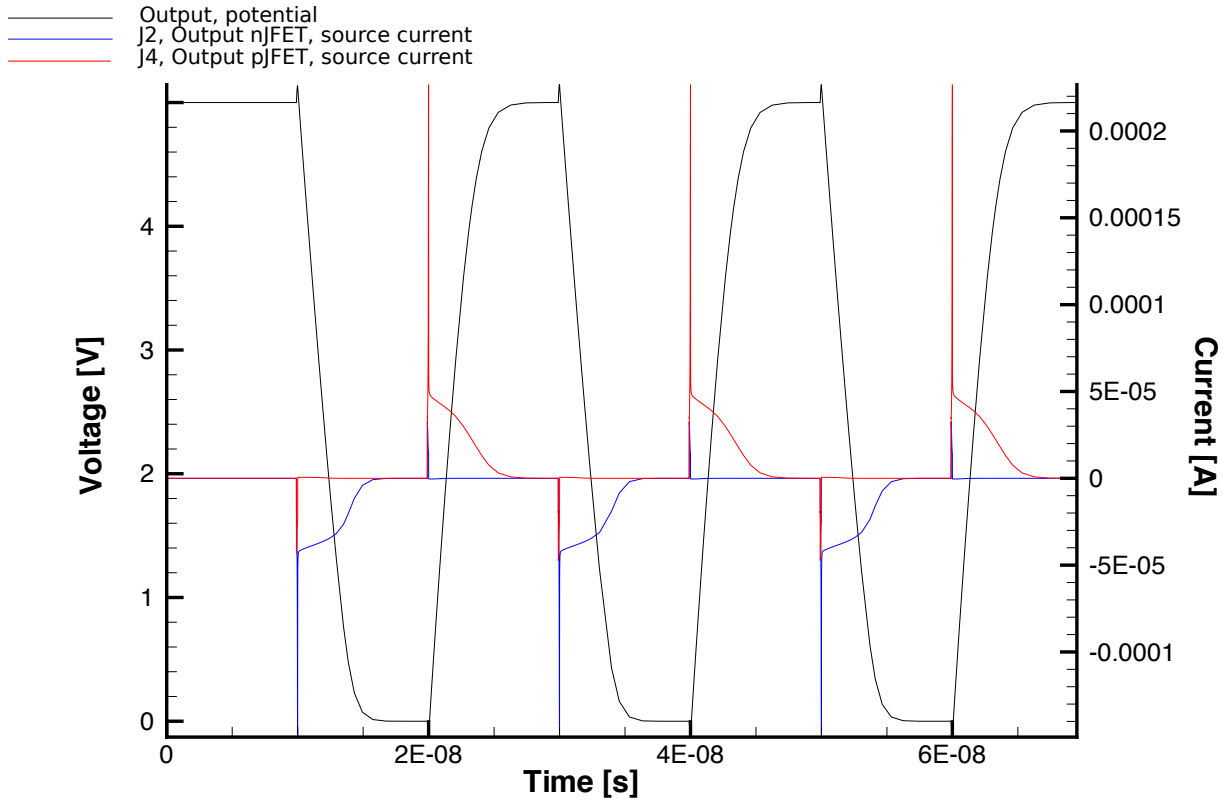

Figure 13. Source current versus time in both output JFETs and the output potential versus time. Based on this image one can clearly see that the output transistors are not simultaneously conductive, i.e., there is no short circuit power dissipation in the simulated CS inverter.

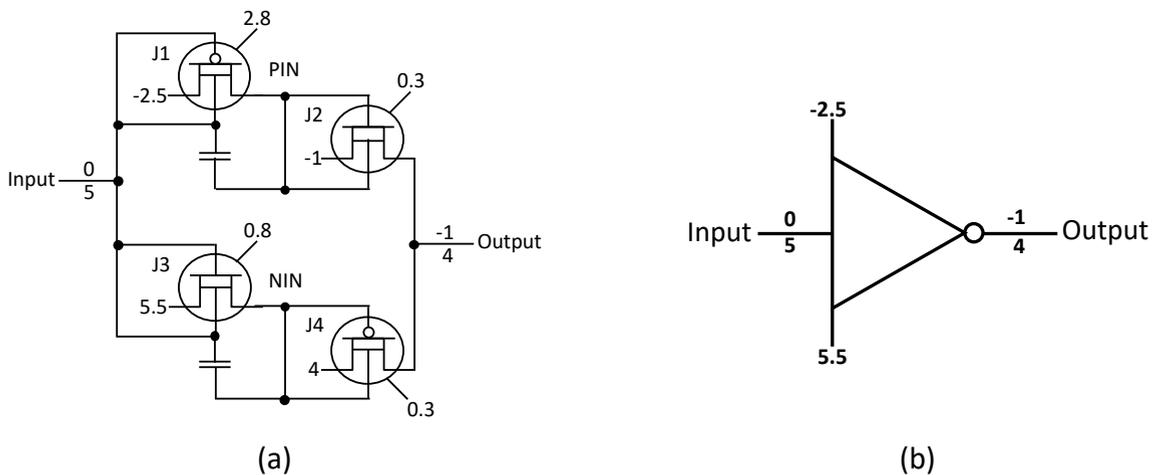

Figure 14. (a) Schematic drawing of a CS inverter configuration wherein the output binary logic levels (-1 and 4 V) are shifted with respect to the input binary logic levels (0 and 5 V). (b) Symbol of the CS inverter configuration presented in figure 14a.

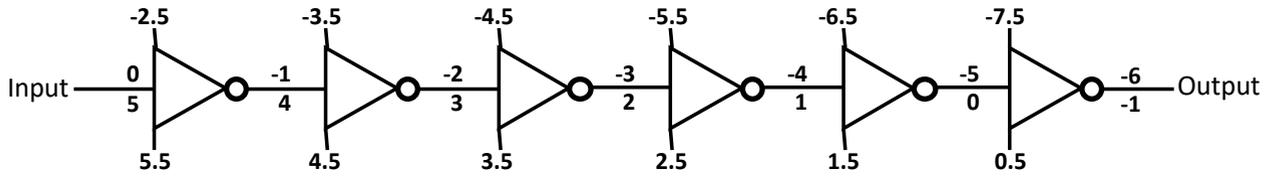

Figure 15. CS inverters are connected in series to shift the binary logic potential pair by -6 V.

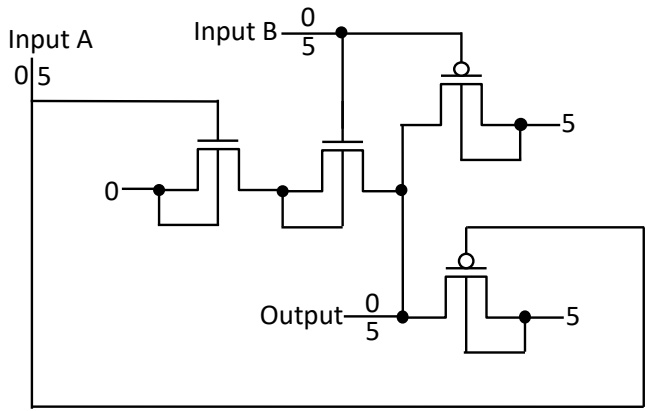

Figure 16a. A CMOS NAND cell with inputs A and B.

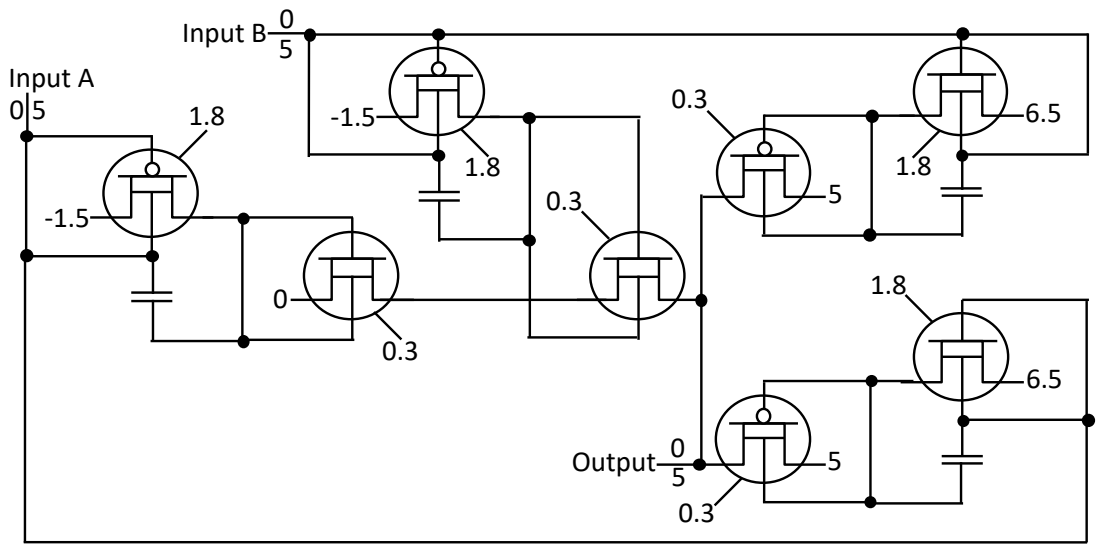

Figure 16b. A CS NAND cell with inputs A and B.

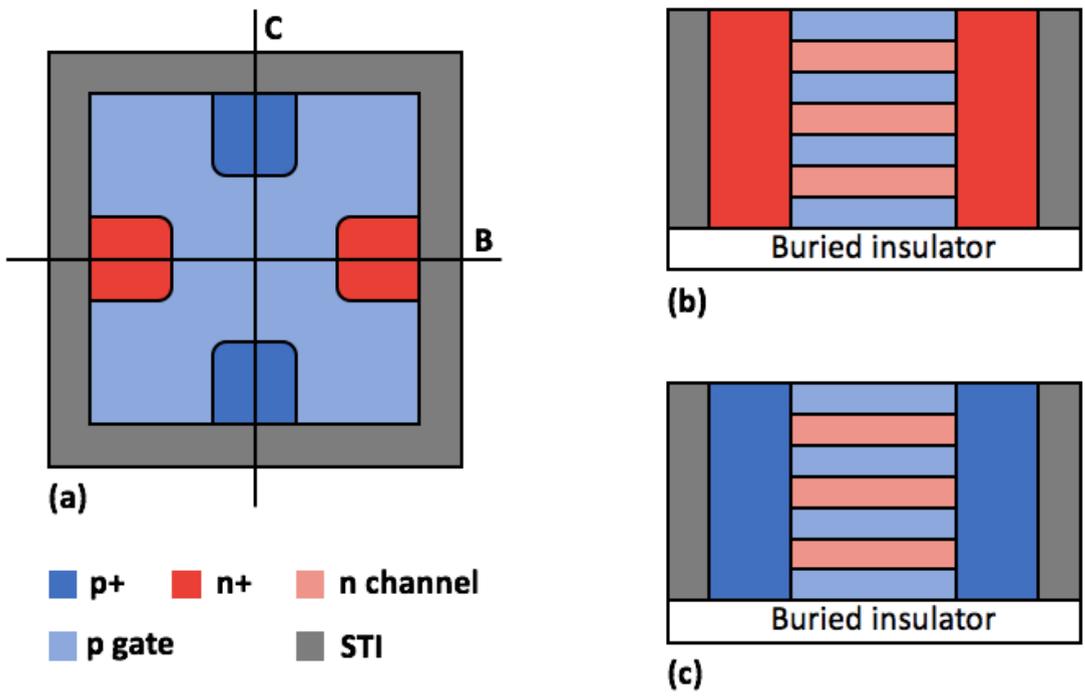

Figure 17. (a) Layout of a multi-channel nJFET architecture. (b) Cross-section along the cutline B. (c) Cross-section along the cutline C.

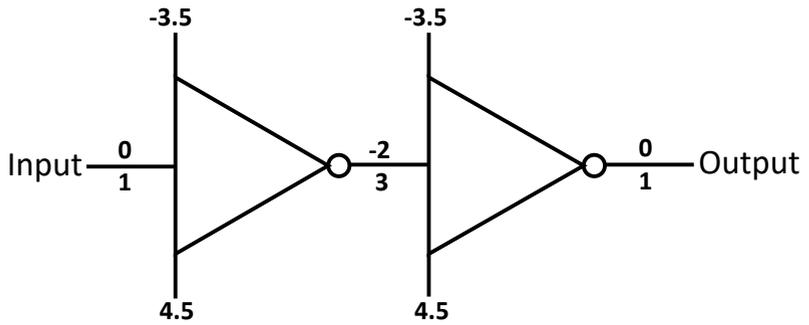

Figure 18. Possible signal booster arrangement comprising two CS inverters.

25